\newcommand{\CLASSINPUTtoptextmargin}{19mm}%
\newcommand{\CLASSINPUTbottomtextmargin}{43mm}%
\newcommand{\CLASSINPUTinnersidemargin}{12.9mm}%
\newcommand{\CLASSINPUToutersidemargin}{12.9mm}%
\documentclass[conference,10pt,a4paper]{IEEEtran}
\usepackage{amsmath}
\usepackage{times}
\usepackage{graphicx}
\usepackage{multirow}
\usepackage[none]{hyphenat}
\usepackage{float}

\usepackage{colortbl}
\usepackage{bm} 
\usepackage{array}
\newcolumntype{L}[1]{>{\vspace{0pt}\raggedright\let\newline\\\arraybackslash\hspace{0pt}}m{#1}}
\newcolumntype{C}[1]{>{\vspace{0pt}\centering\let\newline\\\arraybackslash\hspace{0pt}}m{#1}}
\newcolumntype{R}[1]{>{\vspace{0pt}\raggedleft\let\newline\\\arraybackslash\hspace{0pt}}m{#1}}
\usepackage{tikz}
\usepackage{subcaption}
\usepackage{siunitx}
\usepackage{xcolor}
\usepackage{soul}
\usepackage{graphicx}
\usepackage{textcomp}
\usepackage{caption}
\usepackage{grffile}
\usepackage{psfrag}

\colorlet{soulred}{red!20}
\colorlet{soulblue}{blue!20}
\colorlet{soulgreen}{green!20}
\colorlet{soulmagenta}{magenta!20}
\colorlet{soulviolet}{cyan!20}
\colorlet{soulyellow}{yellow!40}
\usepackage{pgfplots}

\makeatother
\begingroup
\catcode`\_=13
\gdef_#1{\sb{\mathrm{#1}}}
\endgroup
\newcommand\enableuprightsubscripts{\catcode`\_=12\relax}

\makeatletter
\enableuprightsubscripts       
\makeatletter

\def\@maketitle{\newpage
\bgroup\par\addvspace{0.5\baselineskip}\centering%
\ifCLASSOPTIONtechnote
   {\bfseries\large\@IEEEcompsoconly{\sffamily}\@title\par}\vskip 1.3em{\lineskip .5em\@IEEEcompsoconly{\sffamily}\@author
   \@IEEEspecialpapernotice\par{\@IEEEcompsoconly{\vskip 1.5em\relax
   \@IEEEtitleabstractindextextbox{\@IEEEtitleabstractindextext}\par
   \hfill\@IEEEcompsocdiamondline\hfill\hbox{}\par}}}\relax
\else
   \vskip0.2em{\EuMWtitlesize\ifCLASSOPTIONtransmag\bfseries\LARGE\fi\@IEEEcompsoconly{\sffamily}\@IEEEcompsocconfonly{\normalfont\normalsize\vskip 2\@IEEEnormalsizeunitybaselineskip
   \bfseries\Large}\@title\par}\vskip1.0em\par
   \ifCLASSOPTIONconference%
      {\@IEEEspecialpapernotice\mbox{}\vskip\@IEEEauthorblockconfadjspace%
       \mbox{}\hfill\begin{@IEEEauthorhalign}\@author\end{@IEEEauthorhalign}\hfill\mbox{}\par}\relax
   \else
      \ifCLASSOPTIONpeerreviewca
         {\@IEEEcompsoconly{\sffamily}\@IEEEspecialpapernotice\mbox{}\vskip\@IEEEauthorblockconfadjspace%
          \mbox{}\hfill\begin{@IEEEauthorhalign}\@author\end{@IEEEauthorhalign}\hfill\mbox{}\par
          {\@IEEEcompsoconly{\vskip 1.5em\relax
           \@IEEEtitleabstractindextextbox{\@IEEEtitleabstractindextext}\par\hfill
           \@IEEEcompsocdiamondline\hfill\hbox{}\par}}}\relax
      \else
         \ifCLASSOPTIONtransmag
           {\@IEEEspecialpapernotice\mbox{}\vskip\@IEEEauthorblockconfadjspace%
            \mbox{}\hfill\begin{@IEEEauthorhalign}\@author\end{@IEEEauthorhalign}\hfill\mbox{}\par
           {\vspace{0.5\baselineskip}\relax\@IEEEtitleabstractindextextbox{\@IEEEtitleabstractindextext}\vspace{-1\baselineskip}\par}}\relax
         \else
           {\lineskip.5em\@IEEEcompsoconly{\sffamily}\sublargesize\@author\@IEEEspecialpapernotice\par
           {\@IEEEcompsoconly{\vskip 1.5em\relax
            \@IEEEtitleabstractindextextbox{\@IEEEtitleabstractindextext}\par\hfill
            \@IEEEcompsocdiamondline\hfill\hbox{}\par}}}\relax
         \fi
      \fi
   \fi
\fi\par\addvspace{0.0\baselineskip}\egroup}

\def\EuMWtitlesize{\@setfontsize{\EuMWtitlesize}{24}{24pt}}
\def\EuMWauthorsize{\@setfontsize{\EuMWauthorsize}{11}{11pt}}
\def\EuMWaffilsize{\@setfontsize{\EuMWaffilsize}{10}{10pt}}
\def\EuMWcaptionsize{\@setfontsize{\EuMWcaptionsize}{9}{10pt}}
\def\EuMWbibsize{\@setfontsize{\EuMWbibsize}{8}{10pt}}

\def\@IEEEauthorblockNstyle{\EuMWauthorsize\@IEEEcompsocnotconfonly{\sffamily}\@IEEEcompsocconfonly{\large}}
\def\@IEEEauthorblockAstyle{\EuMWaffilsize\@IEEEcompsocnotconfonly{\sffamily}\@IEEEcompsocconfonly{\itshape}\@IEEEcompsocconfonly{\large}}
\def\@IEEEauthordefaulttextstyle{\EuMWauthorsize\@IEEEcompsocnotconfonly{\sffamily}\sublargesize}

\def\thebibliography#1{\section*{\refname}%
    \addcontentsline{toc}{section}{\refname}%
    \EuMWbibsize\@IEEEcompsocconfonly{\small}\vskip 0.3\baselineskip plus 0.1\baselineskip minus 0.1\baselineskip
    \list{\@biblabel{\@arabic\c@enumiv}}%
    {\settowidth\labelwidth{\@biblabel{#1}}%
    \leftmargin\labelwidth
    \advance\leftmargin\labelsep\relax
    \itemsep \IEEEbibitemsep\relax
    \usecounter{enumiv}%
    \let\p@enumiv\@empty
    \renewcommand\theenumiv{\@arabic\c@enumiv}}%
    \let\@IEEElatexbibitem\bibitem%
    \def\bibitem{\@IEEEbibitemprefix\@IEEElatexbibitem}%
\def\newblock{\hskip .11em plus .33em minus .07em}%
\ifCLASSOPTIONtechnote\sloppy\clubpenalty4000\widowpenalty4000\interlinepenalty100%
\else\sloppy\clubpenalty4000\widowpenalty4000\interlinepenalty500\fi%
    \sfcode`\.=1000\relax}

\long\def\@makecaption#1#2{%
\ifx\@captype\@IEEEtablestring%
\par\@IEEEtabletopskipstrut
\else
\@IEEEfigurecaptionsepspace
\fi
\setbox\@tempboxa\hbox{\normalfont\footnotesize {#1.}\nobreakspace\nobreakspace #2}%
\ifdim \wd\@tempboxa >\hsize%
\setbox\@tempboxa\hbox{\normalfont\footnotesize {#1.}\nobreakspace\nobreakspace}%
\parbox[t]{\hsize}{\normalfont\footnotesize\noindent\unhbox\@tempboxa#2}%
\else
\ifCLASSOPTIONconference \hbox to\hsize{\normalfont\footnotesize\hfil\box\@tempboxa\hfil}%
\else \hbox to\hsize{\normalfont\footnotesize\box\@tempboxa\hfil}%
\fi\fi
\ifx\@captype\@IEEEtablestring%
\@IEEEtablecaptionsepspace
\else
\fi}

\newlength\tablecaptiontotableskip
\newlength\figuretocaptionskip
\def\@IEEEfigurecaptionsepspace{\vskip\figuretocaptionskip\relax}%
\def\@IEEEtablecaptionsepspace{\vskip\tablecaptiontotableskip\relax}%

\def\abstract{\normalfont%
\@IEEEabskeysecsize\bfseries\textit{\abstractname}\,\bfseries\textit{---}\,%
\@IEEEgobbleleadPARNLSP}%

\def\IEEEkeywords{\normalfont%
\@IEEEabskeysecsize\bfseries\textit{\IEEEkeywordsname}\,\bfseries\textit{---}\,%
\@IEEEgobbleleadPARNLSP}%
\def\endIEEEkeywords{\relax\vspace{0.67ex}%
\par\if@twocolumn\else\endquotation\fi%
\normalsize\normalfont}%

\DeclareRobustCommand*{\EuMWauthorrefmark}[1]{\raisebox{0pt}[0pt][0pt]{\textsuperscript{\footnotesize{#1}}}}%
\def\@IEEEauthorblockNtopspace{0ex}
\def\@IEEEauthorblockAtopspace{1mm}
\def\tablename{Table}
\def\thetable{\arabic{table}}
\def\IEEEkeywordsname{Keywords}
\def\subsubsection{\@startsection{subsubsection}{3}{\z@}{1.5ex plus 1.5ex minus 0.5ex}%
{0.7ex plus .5ex minus 0ex}{\normalfont\normalsize\itshape}}%
\newlength{\CPheadmatchindent}%
\def\@seccntformat#1{\hbox to\CPheadmatchindent{\csname the#1dis\endcsname}\hskip 0.1em \relax}
\IEEEilabelindentA \parindent
\IEEEilabelindent \IEEEilabelindentA
\IEEEelabelindent \parindent
\IEEEdlabelindent \parindent
\IEEElabelindent \parindent
\makeatother

\begin{document}
\raggedbottom
%
%
\title{Implementation of Real-Time Automotive \\ SAR Imaging}
\author{%
\IEEEauthorblockN{%
Marcel Hoffmann\EuMWauthorrefmark{\#1}, 
Theresa Noegel\EuMWauthorrefmark{\#2},
Christian Schüßler\EuMWauthorrefmark{\#3},
Lars Schwenger\EuMWauthorrefmark{*4},
Peter Gulden\EuMWauthorrefmark{\&5},\\
Dietmar Fey\EuMWauthorrefmark{*6},
Martin Vossiek\EuMWauthorrefmark{\#7}
}
\IEEEauthorblockA{%
\EuMWauthorrefmark{\#}Institute of Microwaves and Photonics (LHFT), Friedrich-Alexander-Universität Erlangen-Nürnberg, Germany\\
\EuMWauthorrefmark{*}Chair of Computer Architecture, Friedrich-Alexander-Universität Erlangen-Nürnberg, Germany\\
\EuMWauthorrefmark{\&}indie Semiconductor, Germany\\
\{\EuMWauthorrefmark{1}marcel.mh.hoffmann, \EuMWauthorrefmark{2}theresa.noegel, \EuMWauthorrefmark{3}christian.schuessler, \EuMWauthorrefmark{4}lars.schwenger, \EuMWauthorrefmark{6}dietmar.fey, \EuMWauthorrefmark{7}martin.vossiek\}@fau.de\\
\EuMWauthorrefmark{5}peter.gulden@indiesemi.com\\
}
}
\maketitle
%
\begin{abstract}
This paper presents measures to reduce the computation time of automotive synthetic aperture radar (SAR) imaging to achieve real-time capability.
For this, the image formation, which is based on the Back-Projection algorithm, was thoroughly analyzed.
Various optimizations were individually tested and analyzed on graphics processing units (GPU).
Apart from the time reduction gained from these measures, the data size needed for processing was also drastically decreased.
With a combination of all measures, a high-resolution SAR image of 30~m $\times$ 30~m that combines 8192 chirps can be reconstructed in less than 30~ms using a standard GPU.
It is thus demonstrated that a real-time implementation of automotive SAR is possible.
\end{abstract}
%
%

\section{Introduction}

For autonomous driving, sensor systems that are capable of generating high-resolution maps of the environment in real time are required.
For this, radars have proven valuable.
To achieve a satisfactory image quality, a high azimuth resolution has to be ensured, which can be realized using large multiple-input multiple-output (MIMO) apertures. \cite{b1}

As the aperture size is limited in vehicles, the azimuth resolution often falls short of requirements. 
It can be improved by generating synthetic apertures that exploit the vehicle’s movement.
These synthetic aperture radar (SAR) approaches typically use sequences of linearly frequency modulated continuous wave signals (FMCW) \cite{b3, b4}. 

Due to constraints including large azimuth variations, inhomogeneous velocities, and curvy trajectories in dynamic automotive scenarios, Back-Projection (BP) is often deployed as a flexible SAR image formation algorithm.
BP optimizes the signal-to-noise ratio (SNR) because it acts as a matched filter and considers the available ego-pose information to the maximum extent without the necessity for simplifying assumptions or equidistant sampling. \cite{b5}

The advantages come at the cost of a high computational complexity. 
For real-time capability, necessary compromises often limit SNR and image quality. 
In \cite{b6}, an embedded, simplified SAR algorithm based on and drastically limited by a preceding target extraction was used. 
The approach in \cite{b7} had to enable uniform sampling by adjusting the pulse repetition frequency. 
In \cite{b8}, Fast Factorized Back-Projection from \cite{b9} was adapted to automotive SAR imaging. 
This reduced the runtime but it could not be easily adapted to changing velocities and the image quality decreased.
Also, because BP is well-suited for parallel processing, computation time can be reduced using a graphics processing unit (GPU) \cite{b10}.

\begin{figure}
	\centering
	
	\begin{subfigure}{8.8cm}
		\begin{minipage}[l]{0.2cm}
			\caption{} \label{camera}
		\end{minipage}
		\hfill
		\begin{minipage}[r]{8.1cm}
			\includegraphics[width=8cm]{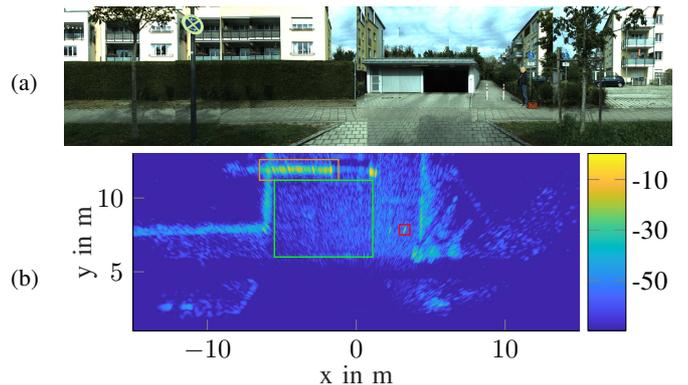}
		\end{minipage}
	\end{subfigure}

	\vspace{0.1cm}	
	
	\begin{subfigure}{8.8cm}
		\begin{minipage}[l]{0.2cm}
			\caption{} \label{fig:holo_ref}
		\end{minipage}
		\hfill
		\begin{minipage}[r]{8.1cm}
			\input{figures/holo_ref_kaesten.tex}
		\end{minipage}
	\end{subfigure}
	
	\caption{a) Camera reference from multiple pictures; b) SAR imaging result in dB normalized to its maximum amplitude with $y$ being the range direction. Red, green, and orange rectangles mark a pole, wall, and yard, respectively.}

\end{figure}

This work aims to process the BP algorithm in real time without compromises in image quality.
For this purpose, several measures for optimizing the algorithm were first evaluated individually and then combined.
With this approach, image formation can be performed within milliseconds, while maintaining the high image quality shown in Fig. \ref{fig:holo_ref}.

\section{Automotive SAR with Back-Projection}

The SAR images were generated using the \SI{77}{\GHz} FMCW MIMO radar system AVR-QDM-110 from Symeo GmbH, an indie Semiconductor company.
It features 12 TX and 16 RX antennas, of which only one TX antenna was used, as the required spatial sampling of $\lambda/4$ is easily fulfilled through vehicle movement.
All radar parameters are listed in Tab. \ref{tab:radar param}.

\begin{table}
	\caption{Chosen FMCW radar parameters.}
	\label{tab:radar param}
	\centering
	\setlength{\tabcolsep}{0.15cm}
	\begin{tabular}{|C{1.5cm}|C{1.5cm}|L{4cm}|}
		\hline
		\textbf{Symbol} & \textbf{Value} & \textbf{Parameter} \\ \hline \hline
		$f_{\text{0}} $ & \SI{76.6}{GHz}& Carrier frequency \\
		$B$ & \SI{931}{MHz}& RF bandwidth\\
		$T_{\text{P}}$ & \SI{102.4}{us}& Chirp duration \\
		$T_{\text{P0}}$ & \SI{106.7}{us}& Pulse repetition interval (PRI) \\
		$N_{\text{m}}$ & $1024$ & Number of FMCW chirps \\
		$N_{tx} \, / \, N_{\text{rx}} $ & $1 \, / \, 16$ & Number of TX / RX antennas \\
		\hline
	\end{tabular}
\end{table}

The radar was roof-mounted to a vehicle looking to the left.
A detailed description of the setup is given in \cite{b4}.

For image formation, the BP algorithm was deployed due to the aforementioned benefits.
Alg.~1 shows the standard BP, acting as a matched filter.
It is based on a correlation of the hypothetically expected signal with the received signal at every reconstruction point (pixel).

\begin{table}[t]
	\centering
	Algorithm 1. \> Standard Back-Projection algorithm.\\[0.7em]
	\setlength{\tabcolsep}{0.1pt}
	\setlength\arrayrulewidth{0.8pt}
	\begin{tabular}{C{0.3cm}L{0.3cm}L{0.3cm}L{0.3cm}L{0.3cm}L{0.3cm}|}
		{\fontsize{6}{9}\selectfont1}& \multicolumn{5}{l}{\textbf{procedure} CALCULATE SAR-IMAGE($\vec{p}$, $\vec{q}_{\text{tx}}$, $\vec{q}_{\text{rx}}$, $\vec{v}$, $s$, $w_{\text{sar}}$)} \\
		\arrayrulecolor{red}
		
		{\fontsize{6}{9}\selectfont2}& & \multicolumn{4}{l}{\sethlcolor{soulred}\hl{\textbf{for all} pixels $p$ \textbf{do}}} \\
		
		\arrayrulecolor{blue}
		{\fontsize{6}{9}\selectfont3}& & & \multicolumn{3}{l}{\sethlcolor{soulblue}\hl{\textbf{for all} measurements $m$ \textbf{do}}} \\
		{\fontsize{6}{9}\selectfont4}& & & & \multicolumn{2}{l}{$d_{\text{tx}}(p,m) \gets || \vec{p}(p) - \vec{q}_{\text{tx}}(m)||_2$} \\
		
		{\fontsize{6}{9}\selectfont5}& & & & \multicolumn{2}{l}{\sethlcolor{soulgreen}\hl{$v_{\text{tx}}(p,m) \gets \langle \vec{p}(p) - \vec{q}_{\text{tx}}(m), \vec{v}(m) \rangle / d_{\text{tx}}(p,m)$} }\\

		{\fontsize{6}{9}\selectfont6}& & & & \multicolumn{2}{l}{\sethlcolor{soulblue}\hl{\textbf{for all} rx antennas $n_{\text{rx}}$ \textbf{do}}} \\
		
		
		{\fontsize{6}{9}\selectfont7}& & & & & \multicolumn{1}{l}{$d_{\text{rx}}(p,m,n_{\text{rx}}) \gets || \vec{p}(p) - \vec{q}_{\text{rx}}(m,n_{\text{rx}}) ||_2$} \\
		\arrayrulecolor{green}
		
		{\fontsize{6}{9}\selectfont8}& & & & & \multicolumn{1}{l}{\sethlcolor{soulgreen}\hl{$v_{\text{rx}}(p,m,n_{\text{rx}}) \gets \langle \vec{p}(p) - \vec{q}_{\text{rx}}(m, n_{\text{rx}}), \vec{v}(m) \rangle / d_{\text{rx}}(p,m,n_{\text{rx}})$} }\\		
		
		{\fontsize{6}{9}\selectfont9}& & & & & \multicolumn{1}{l}{\sethlcolor{soulgreen}\hl{$v(p,m,n_{\text{rx}}) \gets v_{\text{tx}}(p,m) + v_{\text{rx}}(p,m,n_{\text{rx}})$}} \\	
				
		{\fontsize{6}{9}\selectfont10}& & & & & \multicolumn{1}{l}{$\tau(p,m,n_{\text{rx}}) \gets ( d_{\text{tx}}(p,m) + d_{\text{rx}}(p,m,n_{\text{rx}})) / c$} \\
		
		{\fontsize{6}{9}\selectfont11}& & & & & \multicolumn{1}{l}{$f_{\text{hyp}}(p,mn_{\text{rx}}) \gets \mu \cdot \tau(p,m,n_{\text{rx}}) + f_{\text{0}} \cdot v(p,m,n_{\text{rx}}) /c $} \\
		{\fontsize{6}{9}\selectfont12}& & & & & \multicolumn{1}{l}{$s_{\text{hyp}}(p,m,n_{\text{rx}}) \gets \exp(-\text{j} \cdot 2 \pi \cdot \tau(p,m,n_{\text{rx}}))$} \\
		{\fontsize{6}{9}\selectfont13}& & & & & \multicolumn{1}{l}{$P(p,m,n_{\text{rx}}) \gets s_{\text{hyp}}(p,m,n_{\text{rx}}) \cdot$ \sethlcolor{soulmagenta}\hl{$s(f_{\text{hyp}},m,n_{\text{rx}}))$} $\cdot $\sethlcolor{soulviolet}\hl{$w_{sar}$}} \\
		{\fontsize{6}{9}\selectfont14}& & \multicolumn{4}{l}{\textbf{for all} pixels $p$ \textbf{do}} \\
		{\fontsize{6}{9}\selectfont15}& & & \multicolumn{3}{l}{$P(p) \gets \sum_{m}^{} \sum_{n_{\text{rx}}}^{}  P(p,m,n_{\text{rx}})$} \\
		
	\end{tabular}
\end{table}

In lines 4 and 7, for every chirp $m$ the 2D distance $d_{tx}$ and $d_{rx}$ is calculated between each pixel $p$ at the position $\vec{p}$ and the TX antenna at the position $\vec{q}_{\text{tx}}$ and the RX antenna $n_{\text{rx}}$ at the position $\vec{q}_{\text{rx}}$, respectively.
In this context, $||\cdot ||_2$ denotes the Euclidean norm.
The Doppler shift caused by the radial component of the vehicle's velocity $\vec{v}$ is considered pixelwise in lines 5 and 8 for the TX and RX perspective, respectively, and added in line 9.
Here, $\langle \cdot , \cdot \rangle $ represents the scalar product. 
In line 10, the two-way propagation time $\tau$ to a pixel is calculated for every TX-RX and chirp combination.
Based on $\tau$ and $v$, the hypothetical beat frequency $f_{\text{hyp}}$ is determined in line 11.
For this, the speed of light $c$ and chirp rate $\mu = B / T_{\text{P}}$ are used.
Similarly, a hypothetical beat signal $s_{\text{hyp}}$ can be composed in line 12.
In line 13, $s_{\text{hyp}}$ is correlated with the complex value of the measured beat signal $s$ that is interpolated at $f_{hyp}$.
This is weighted with a window function $w_{sar}$, which is typically a Hann function.
It aims at suppressing sidelobes and is theoretically individual for every pixel and its effective aperture.
Finally, all complex-valued contributions are coherently summed to form the SAR image $P$.

For an initial test, a \qtyproduct{30 x 30}{\m} area was reconstructed with a grid resolution of \SI{2.5}{\cm} resulting in \numproduct{1201 x 1201} pixels.
Using an Intel i7-9850H vPro CPU, the required computation time of this reference approach was \SI{822.1}{\s}.
A \qtyproduct{30 x 12}{\m} section of the SAR image is presented in Fig. \ref{fig:holo_ref}, with the amplitude being scaled in dB using $20\log_{10}$.
A precise analysis of the immediate high image quality can also be found in \cite{b4}.
For further evaluation, the right pole, the garage wall, and the yard in front of the garage are highlighted by red, green, and orange rectangles, respectively.
For this reference approach, the average SNR of the three regions and the computation time for the CPU are also summarized in row 1 of Tab. \ref{tab:results}.

%
%

\section{Measures for Reducing Computing Time}

To achieve real-time capability, the computation time must be less than the measurement time defined by $N_{\text{m}} \cdot T_{\text{P0}}$, which is \SI{109.3}{\ms} in our case.
For a reduction of the processing time, we propose six measures.
As a metric for the image quality, we evaluated the SNR reduction of the three areas highlighted in Fig. \ref{fig:holo_ref}.
All results are documented in Tab. \ref{tab:results}.

\subsection{Implementation on a GPU}

As a first measure, the algorithm presented in Alg. 1 was implemented on two different systems using GPUs and a single-precision floating point format.
These were:

\begin{itemize}
\item GPU 1: AMD Ryzen 3 PRO 1200 processor with an NVIDIA GeForce GTX 1080 Ti with \num{3584} Cores, \num{11.340} TFLOPS, and \SI{11}{\giga\byte} GDDR5X RAM
\item GPU 2: Intel Core i7-9850H vPro CPU with an NVIDIA Quadro RTX 3000 Mobile with \num{1920} cores, \num{5.299} TFLOPS, and \SI{6}{\giga\byte} GDDR6 RAM
\end{itemize}

We used Python with pyCUDA for the implementation.
Because the entire algorithm was executed on the GPU and implemented in CUDA, the reduced performance of Python compared to languages like C++ is irrelevant.
The algorithm on the GPU was parallelized along the pixels and constant memory was utilized for variables like $\vec{p}$ and $v$.
Because these variables are loaded by all threads, constant memory offers a faster access and improves the runtime.

We measured the loading time to and from the GPU and the calculation runtime through CUDA events and executed each version 20 times.
Tab. \ref{tab:results} shows the measured times in milliseconds for the reference approach in row 1.
As highlighted in yellow, GPU 2 showed a much faster loading speed because GPU 1 was limited by its slower CPU.
However, the actual calculation can be executed faster on GPU 1 as it features considerably more cores.

\begin{table*}[t]
	\caption{SNR and measured times for the tested measures. Colors match the highlights in the standard and optimized BP.}
	\centering
	\setlength{\tabcolsep}{0.15cm}
	\renewcommand{\arraystretch}{1}
	\begin{tabular}{|C{0.2cm}||L{1.2cm}||L{1cm}|R{1cm}||L{0.8cm}|R{1cm}|R{0.7cm}|R{1cm}|R{0.7cm}|R{1cm}|R{0.7cm}|}
		\hline
		\multirow{2}{*}{\textbf{\#}} &\multirow{2}{*}{\textbf{Measure}} & \multicolumn{2}{c||}{\multirow{2}{*}{\textbf{$\bm{\Delta SNR}$ in dB}}} & \multirow{2}{*}{\textbf{Time}} & \multicolumn{2}{c|}{CPU} & \multicolumn{2}{c|}{GPU 1 (1080Ti)} & \multicolumn{2}{c|}{GPU 2 (3000)} \\ 
		\cline{6-11}
		& &\multicolumn{2}{l||}{} & & $t$ in s & \% & $t$ in ms & \% & $t$ in ms & \% \\ \hline \hline
		& & Garage & $54.4$ & Load & - & - & \sethlcolor{soulyellow}\hl{$1985.3$} & $100 $ & \sethlcolor{soulyellow}\hl{$612.2$} & $100 $\\ 
		$1$ & \textbf{Ref.} & Pole & $63.5$ & BP & - & - & \sethlcolor{soulyellow}\hl{$555.6$} & $100 $ & \sethlcolor{soulyellow}\hl{$1036.9$} & $100$ \\ 
		& & Yard & $17.5$ & Total & $822.1$ & $100 $ & $2542.4$ & $100 $ & $1650.3$ & $100 $ \\ \hline
		
		& & Garage & \sethlcolor{soulviolet}\hl{$+0.6$} & Load & - & - & $1496.9$ & \sethlcolor{soulviolet}\hl{$75.4$} & $108.1$ & \sethlcolor{soulviolet}\hl{$17.7$} \\ 
		$2$ & $\bm{w_\text{sar}}$ & Pole & \sethlcolor{soulviolet}\hl{$+0.6$} & BP & - & - & $428.1$ & $77.0 $& $781.0$ &$75.3 $ \\ 
		& & Yard & \sethlcolor{soulviolet}\hl{$0$} & Total & $761.2$ &$92.6 $ & $1926.2$ &$75.8 $ & $890.2$ & $53.9 $\\ \hline
		
		& & Garage & $0$ & Load & - & - & $1980.6$ &$99.8 $ & $612.1$ &$100 $ \\ 
		$3$ & \textbf{Opt.} & Pole & $0$ & BP & - & - & $465.2$ &$83.7 $ & $731.2$ &$70.5 $ \\ 
		& & Yard & $0$ & Total & $766.8$ &$93.30 $ & $2447.0$ &$96.2 $ & $1351.6$ &$81.9 $ \\ \hline
		
		& & Garage & $0$ & Load & - & - & $1982.2$ &$99.8 $ & $611.8$ &$99.9 $ \\ 
		$4$ & \textbf{Doppler} & Pole & $0$ & BP & - & - & $353.8$ & \sethlcolor{soulgreen}\hl{$63.7$} & $673.6$ & \sethlcolor{soulgreen}\hl{$65.0$} \\ 
		& & Yard & $0$ & Total & $768.7$ &$93.5 $ & $2337.2$ &$91.9 $ & $1286.6$ &$78.0 $ \\ \hline

		& & Garage& $-0.5$ & Load & - & - & $1557.7$ &$78.5 $ & $167.3$ &$27.3 $ \\ 
		$5$ & \textbf{Polar} & Pole & $-0.9$ & BP & - & - & $66.0$ & \sethlcolor{soulred}\hl{$11.9 $} & $107.4$ & \sethlcolor{soulred}\hl{$10.4 $} \\ 
		& & Yard & $0$ & Total & $123.8$ & \sethlcolor{soulred}\hl{$15.1 $} & $1624.1$ &$63.9 $ & $275.0$ &$16.7 $ \\ \hline

		& & Garage & $-1.0$ & Load & - & - & $1189.5$ & $59.9 $ & $565.6$ &$92.4 $ \\ 
		$6$ & \textbf{RX Ant.} & Pole & $-4.8$ & BP & - & - & $287.3$ & \sethlcolor{soulblue}\hl{$51.7 $} & $545.2$ & \sethlcolor{soulblue}\hl{$52.6 $} \\ 
		& & Yard & $-3.9$ & Total & $415.6$ & \sethlcolor{soulblue}\hl{$50.6 $} & $1478.0$ &$58.1 $ & $1112.0$ &$67.4 $ \\ \hline

		& & Garage & $+0.8$& Load & - & - & \sethlcolor{soulyellow}\hl{$315.7$} &$15.9 $ & \sethlcolor{soulyellow}\hl{$28.1$} &$4.6 $ \\ 
		$7$ & \textbf{Comb.} & Pole & $-1.5$ & BP & - & - & $16.1$ &$2.9 $ & $25.1$ &$2.4 $ \\ 
		& & Yard & $-2.0$ & Total & $37.1$ &$4.5 $ & $332.0$ &$13.1 $ & $53.4$ &$3.2 $ \\ \hline
		
		& \textbf{Comb.} & Garage & - & Load & - & - & \sethlcolor{soulyellow}\hl{$3.0$} &\sethlcolor{soulyellow}\hl{$0.2 $} & \sethlcolor{soulyellow}\hl{$2.9$} & \sethlcolor{soulyellow}\hl{$0.5 $} \\ 
		$8$ & \textbf{Pinned} & Pole & - & BP & - & - & $15.5$ &$2.8 $ & $25.1$ &$2.4 $ \\ 
		& \textbf{Memory} & Yard & - & Total & - & - & \sethlcolor{soulyellow}\hl{$18.8$} & \sethlcolor{soulyellow}\hl{$0.7 $} & \sethlcolor{soulyellow}\hl{$28.4$} & \sethlcolor{soulyellow}\hl{$1.7 $} \\ \hline
		
	\end{tabular}
	\label{tab:results}
\end{table*}

\subsection{Adaption of the Window Function}

In Alg. 1, the window function $w_{sar}$ used a theoretically correct, non-equidistantly sampled weight for every pixel and chirp combination.
Therefore, it was of the dimensions \numproduct{1201 x 1201 x 1024}, had a size of \SI{5.91}{\giga\byte}, and made up \SI{98}{\percent} of the complete size of \SI{6.04}{\giga\byte} loaded to the GPU.
This precision is not required, since the amplitude of target areas is primarily influenced by the coherent superposition of phase values and not by the signals' amplitudes.
Also, for the weight values, the velocity can be considered constant during the measurement.
Consequently, an equidistantly sampled window function that is only dependent on the chirp number is sufficient and reduces the dimension drastically to a vector of size \num{1024} with merely \SI{4.10}{\kilo\byte}.
With this measure, the loading time for GPU 1 and 2 was reduced to \SI{75.4}{\percent} and \SI{17.7}{\percent}, respectively (blue values in Tab. \ref{tab:results}).
Also, the calculation time was reduced by the simplified multiplication in line 13 of Alg.~1.

Moreover, an analysis of the noise floor has shown that a Kaiser window sufficiently reduces the sidelobes and slightly increases the resolution and SNR of targets, which is also highlighted in light blue in Tab. \ref{tab:results}.

\subsection{Mathematical Optimizations}

This measure focuses on mathematical optimizations to substitute unnecessary calculations as constants that can be performed beforehand.
Also, instead of $f_{hyp}$, only the index of the signal vector $s$ is required for linear interpolation.
As the corresponding beat frequency axis is equidistantly spaced, the interpolation in line 13 in Alg. 1 can be simplified as shown in pink in line 10 in the optimized BP algorithm in Alg. 2.

\subsection{Simplification of the Doppler Component}

In lines 5, 8, and 9 in Alg. 1, the velocity component is elaborately calculated for the precise determination of $f_{hyp}$.
Because the Doppler frequency can lie in the range of several kHz, it cannot be neglected for short distances, where the beat frequency is only a few hundred kHz.
However, equivalent to the window function, the velocity can be considered constant within a chirp sequence, because the variation only accounts for deviations well below \SI{100}{\Hz}. 
Therefore, the radial component for every pixel $p$ can be calculated in advance based on the average vehicle velocity. 
The green simplification in Alg. 2. had no impact on the SNR in Tab. \ref{tab:results}, but reduced the runtime by more than \SI{35}{\percent} compared to the reference.

\subsection{Reduction of the Reconstruction Points}

\begin{figure}[tb]
	\centering
	\input{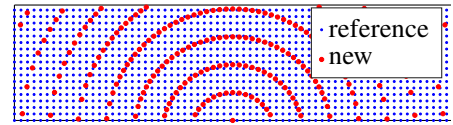}
	\caption{Reference and proposed reconstruction grid.}
	\label{fig:skizze_polar}
\end{figure}

The reference implementation was based on a Cartesian reconstruction grid with a resolution of \SI{2.5}{\cm}.
This does not consider the polar resolution characteristics of the radar and is too fine, especially for larger distances.
Therefore, we propose a new grid in Fig. \ref{fig:skizze_polar} that originates in the center of the synthetic aperture.
Its range and azimuth resolution depends on the point spread function (PSF) of the radar, which is described in detail in \cite{b4}.
To not affect or reduce the image resolution, the grid resolution must be a certain factor finer than both the range and azimuth resolution.
The factor must be larger than 2 to correctly depict point targets.
To avoid too many grid points, a factor of 2.5 has proven to be a good compromise.
With this approach, the number of pixels can be reduced to \num{165061}, which is \SI{11.5}{\percent} compared to the reference.
This is in almost perfect agreement with the runtime improvements to \SI{11.9}{\percent} and \SI{10.4}{\percent} highlighted in red in Tab. \ref{tab:results}.

\subsection{Reduction of Aperture Points}

The reference setup used 16 RX and \num{1024} chirps, which led to a severe spatial oversampling but not a significant image improvement.
Due to a non-perfect calibration of the MIMO array, a phase-coherent superposition of a larger number of chirps is more effective than using a larger number of RX antennas with slight phase deviations. 
We therefore reduced the number of antennas until the image quality decreased.
The best compromise was achieved using 8 RX antennas, which reduced the number of computational operations and radar data to \SI{50}{\percent}. 
This is also reflected in Tab. \ref{tab:results} in the values highlighted in purple.
As the optimization of $w_{sar}$ had not yet been included in this test, the loading time was dominated by the window function and the reduction was less significant.
By reducing the number of RX antennas by a factor of two, the amplitude of several image areas was decreased by up to \SI{6}{\dB}.
Due to the imperfect calibration, this is typically lower than \SI{6}{\dB} as can be seen by the SNR values in Tab. \ref{tab:results}.

\section{Performance Result With Combination of All Measures}

After an individual analysis of all measures, a combined version was created.
The optimized algorithm incorporating all aforementioned measures is given in Alg. 2.
The color highlights match the measures in Alg. 1 and in Tab. \ref{tab:results}.
With this approach, the total time was reduced to \SI{332.0}{\ms} and \SI{53.4}{\ms} or \SI{13.1}{\percent} and \SI{3.2}{\percent} of the reference time for GPU 1 and 2, respectively.
Also, the data size was reduced from over \SI{6}{\giga\byte} to merely \SI{30.5}{\mega\byte}, which is \SI{0.5}{\percent} of the starting value.

However, the loading time was still quite high, which is highlighted in yellow.
This was especially relevant for GPU~1, as it is limited by its slower CPU.
This drawback can be circumvented by pinning the memory to the GPU, removing most CPU overhead in the copying process.
Then, both GPUs show similar host-to-device copying times (row 8 in Tab.~\ref{tab:results}).
Even though GPU 1 was faster due to its higher number of cores, GPU 2 showed a strong enhancement, showcasing that smaller and slower GPUs can also handle the task through our improvements.
With a total computation time of \SI{18.8}{\ms} and \SI{28.4}{\ms}, this approach can be considered real-time capable, as this was much smaller than the radar's measurement time of \SI{109.3}{\ms}.
Furthermore, constraints on global GPU memory as a design criterion have been almost entirely removed, which is a major asset for embedded systems and power consumption. 

\begin{figure}
	\centering
	
	\begin{subfigure}{8.8cm}
		\begin{minipage}[l]{0.2cm}
			\caption{} \label{holo_kombi}
		\end{minipage}
		\hfill
		\begin{minipage}[r]{8.1cm}
%
%
\begin{tikzpicture}

\begin{axis}[%
scale = 0.58, 
width=3.987in,
height=1.993in*12/15,
at={(0.669in,1.267in)},
scale only axis,
point meta min=100.526993340352,
point meta max=170.526993340352,
axis on top,
xmin=-15,
xmax=15,
xtick=\empty,
xlabel style={font=\color{white!15!black}},
ymin=1,
ymax=13,
ylabel style={font=\color{white!15!black}, yshift = -0.6cm},
ylabel={y in m},
axis background/.style={fill=white},
title style={font=\bfseries},
colormap={mymap}{[1pt] rgb(0pt)=(0.2422,0.1504,0.6603); rgb(1pt)=(0.2444,0.1534,0.6728); rgb(2pt)=(0.2464,0.1569,0.6847); rgb(3pt)=(0.2484,0.1607,0.6961); rgb(4pt)=(0.2503,0.1648,0.7071); rgb(5pt)=(0.2522,0.1689,0.7179); rgb(6pt)=(0.254,0.1732,0.7286); rgb(7pt)=(0.2558,0.1773,0.7393); rgb(8pt)=(0.2576,0.1814,0.7501); rgb(9pt)=(0.2594,0.1854,0.761); rgb(11pt)=(0.2628,0.1932,0.7828); rgb(12pt)=(0.2645,0.1972,0.7937); rgb(13pt)=(0.2661,0.2011,0.8043); rgb(14pt)=(0.2676,0.2052,0.8148); rgb(15pt)=(0.2691,0.2094,0.8249); rgb(16pt)=(0.2704,0.2138,0.8346); rgb(17pt)=(0.2717,0.2184,0.8439); rgb(18pt)=(0.2729,0.2231,0.8528); rgb(19pt)=(0.274,0.228,0.8612); rgb(20pt)=(0.2749,0.233,0.8692); rgb(21pt)=(0.2758,0.2382,0.8767); rgb(22pt)=(0.2766,0.2435,0.884); rgb(23pt)=(0.2774,0.2489,0.8908); rgb(24pt)=(0.2781,0.2543,0.8973); rgb(25pt)=(0.2788,0.2598,0.9035); rgb(26pt)=(0.2794,0.2653,0.9094); rgb(27pt)=(0.2798,0.2708,0.915); rgb(28pt)=(0.2802,0.2764,0.9204); rgb(29pt)=(0.2806,0.2819,0.9255); rgb(30pt)=(0.2809,0.2875,0.9305); rgb(31pt)=(0.2811,0.293,0.9352); rgb(32pt)=(0.2813,0.2985,0.9397); rgb(33pt)=(0.2814,0.304,0.9441); rgb(34pt)=(0.2814,0.3095,0.9483); rgb(35pt)=(0.2813,0.315,0.9524); rgb(36pt)=(0.2811,0.3204,0.9563); rgb(37pt)=(0.2809,0.3259,0.96); rgb(38pt)=(0.2807,0.3313,0.9636); rgb(39pt)=(0.2803,0.3367,0.967); rgb(40pt)=(0.2798,0.3421,0.9702); rgb(41pt)=(0.2791,0.3475,0.9733); rgb(42pt)=(0.2784,0.3529,0.9763); rgb(43pt)=(0.2776,0.3583,0.9791); rgb(44pt)=(0.2766,0.3638,0.9817); rgb(45pt)=(0.2754,0.3693,0.984); rgb(46pt)=(0.2741,0.3748,0.9862); rgb(47pt)=(0.2726,0.3804,0.9881); rgb(48pt)=(0.271,0.386,0.9898); rgb(49pt)=(0.2691,0.3916,0.9912); rgb(50pt)=(0.267,0.3973,0.9924); rgb(51pt)=(0.2647,0.403,0.9935); rgb(52pt)=(0.2621,0.4088,0.9946); rgb(53pt)=(0.2591,0.4145,0.9955); rgb(54pt)=(0.2556,0.4203,0.9965); rgb(55pt)=(0.2517,0.4261,0.9974); rgb(56pt)=(0.2473,0.4319,0.9983); rgb(57pt)=(0.2424,0.4378,0.9991); rgb(58pt)=(0.2369,0.4437,0.9996); rgb(59pt)=(0.2311,0.4497,0.9995); rgb(60pt)=(0.225,0.4559,0.9985); rgb(61pt)=(0.2189,0.462,0.9968); rgb(62pt)=(0.2128,0.4682,0.9948); rgb(63pt)=(0.2066,0.4743,0.9926); rgb(64pt)=(0.2006,0.4803,0.9906); rgb(65pt)=(0.195,0.4861,0.9887); rgb(66pt)=(0.1903,0.4919,0.9867); rgb(67pt)=(0.1869,0.4975,0.9844); rgb(68pt)=(0.1847,0.503,0.9819); rgb(69pt)=(0.1831,0.5084,0.9793); rgb(70pt)=(0.1818,0.5138,0.9766); rgb(71pt)=(0.1806,0.5191,0.9738); rgb(72pt)=(0.1795,0.5244,0.9709); rgb(73pt)=(0.1785,0.5296,0.9677); rgb(74pt)=(0.1778,0.5349,0.9641); rgb(75pt)=(0.1773,0.5401,0.9602); rgb(76pt)=(0.1768,0.5452,0.956); rgb(77pt)=(0.1764,0.5504,0.9516); rgb(78pt)=(0.1755,0.5554,0.9473); rgb(79pt)=(0.174,0.5605,0.9432); rgb(80pt)=(0.1716,0.5655,0.9393); rgb(81pt)=(0.1686,0.5705,0.9357); rgb(82pt)=(0.1649,0.5755,0.9323); rgb(83pt)=(0.161,0.5805,0.9289); rgb(84pt)=(0.1573,0.5854,0.9254); rgb(85pt)=(0.154,0.5902,0.9218); rgb(86pt)=(0.1513,0.595,0.9182); rgb(87pt)=(0.1492,0.5997,0.9147); rgb(88pt)=(0.1475,0.6043,0.9113); rgb(89pt)=(0.1461,0.6089,0.908); rgb(90pt)=(0.1446,0.6135,0.905); rgb(91pt)=(0.1429,0.618,0.9022); rgb(92pt)=(0.1408,0.6226,0.8998); rgb(93pt)=(0.1383,0.6272,0.8975); rgb(94pt)=(0.1354,0.6317,0.8953); rgb(95pt)=(0.1321,0.6363,0.8932); rgb(96pt)=(0.1288,0.6408,0.891); rgb(97pt)=(0.1253,0.6453,0.8887); rgb(98pt)=(0.1219,0.6497,0.8862); rgb(99pt)=(0.1185,0.6541,0.8834); rgb(100pt)=(0.1152,0.6584,0.8804); rgb(101pt)=(0.1119,0.6627,0.877); rgb(102pt)=(0.1085,0.6669,0.8734); rgb(103pt)=(0.1048,0.671,0.8695); rgb(104pt)=(0.1009,0.675,0.8653); rgb(105pt)=(0.0964,0.6789,0.8609); rgb(106pt)=(0.0914,0.6828,0.8562); rgb(107pt)=(0.0855,0.6865,0.8513); rgb(108pt)=(0.0789,0.6902,0.8462); rgb(109pt)=(0.0713,0.6938,0.8409); rgb(110pt)=(0.0628,0.6972,0.8355); rgb(111pt)=(0.0535,0.7006,0.8299); rgb(112pt)=(0.0433,0.7039,0.8242); rgb(113pt)=(0.0328,0.7071,0.8183); rgb(114pt)=(0.0234,0.7103,0.8124); rgb(115pt)=(0.0155,0.7133,0.8064); rgb(116pt)=(0.0091,0.7163,0.8003); rgb(117pt)=(0.0046,0.7192,0.7941); rgb(118pt)=(0.0019,0.722,0.7878); rgb(119pt)=(0.0009,0.7248,0.7815); rgb(120pt)=(0.0018,0.7275,0.7752); rgb(121pt)=(0.0046,0.7301,0.7688); rgb(122pt)=(0.0094,0.7327,0.7623); rgb(123pt)=(0.0162,0.7352,0.7558); rgb(124pt)=(0.0253,0.7376,0.7492); rgb(125pt)=(0.0369,0.74,0.7426); rgb(126pt)=(0.0504,0.7423,0.7359); rgb(127pt)=(0.0638,0.7446,0.7292); rgb(128pt)=(0.077,0.7468,0.7224); rgb(129pt)=(0.0899,0.7489,0.7156); rgb(130pt)=(0.1023,0.751,0.7088); rgb(131pt)=(0.1141,0.7531,0.7019); rgb(132pt)=(0.1252,0.7552,0.695); rgb(133pt)=(0.1354,0.7572,0.6881); rgb(134pt)=(0.1448,0.7593,0.6812); rgb(135pt)=(0.1532,0.7614,0.6741); rgb(136pt)=(0.1609,0.7635,0.6671); rgb(137pt)=(0.1678,0.7656,0.6599); rgb(138pt)=(0.1741,0.7678,0.6527); rgb(139pt)=(0.1799,0.7699,0.6454); rgb(140pt)=(0.1853,0.7721,0.6379); rgb(141pt)=(0.1905,0.7743,0.6303); rgb(142pt)=(0.1954,0.7765,0.6225); rgb(143pt)=(0.2003,0.7787,0.6146); rgb(144pt)=(0.2061,0.7808,0.6065); rgb(145pt)=(0.2118,0.7828,0.5983); rgb(146pt)=(0.2178,0.7849,0.5899); rgb(147pt)=(0.2244,0.7869,0.5813); rgb(148pt)=(0.2318,0.7887,0.5725); rgb(149pt)=(0.2401,0.7905,0.5636); rgb(150pt)=(0.2491,0.7922,0.5546); rgb(151pt)=(0.2589,0.7937,0.5454); rgb(152pt)=(0.2695,0.7951,0.536); rgb(153pt)=(0.2809,0.7964,0.5266); rgb(154pt)=(0.2929,0.7975,0.517); rgb(155pt)=(0.3052,0.7985,0.5074); rgb(156pt)=(0.3176,0.7994,0.4975); rgb(157pt)=(0.3301,0.8002,0.4876); rgb(158pt)=(0.3424,0.8009,0.4774); rgb(159pt)=(0.3548,0.8016,0.4669); rgb(160pt)=(0.3671,0.8021,0.4563); rgb(161pt)=(0.3795,0.8026,0.4454); rgb(162pt)=(0.3921,0.8029,0.4344); rgb(163pt)=(0.405,0.8031,0.4233); rgb(164pt)=(0.4184,0.803,0.4122); rgb(165pt)=(0.4322,0.8028,0.4013); rgb(166pt)=(0.4463,0.8024,0.3904); rgb(167pt)=(0.4608,0.8018,0.3797); rgb(168pt)=(0.4753,0.8011,0.3691); rgb(169pt)=(0.4899,0.8002,0.3586); rgb(170pt)=(0.5044,0.7993,0.348); rgb(171pt)=(0.5187,0.7982,0.3374); rgb(172pt)=(0.5329,0.797,0.3267); rgb(173pt)=(0.547,0.7957,0.3159); rgb(175pt)=(0.5748,0.7929,0.2941); rgb(176pt)=(0.5886,0.7913,0.2833); rgb(177pt)=(0.6024,0.7896,0.2726); rgb(178pt)=(0.6161,0.7878,0.2622); rgb(179pt)=(0.6297,0.7859,0.2521); rgb(180pt)=(0.6433,0.7839,0.2423); rgb(181pt)=(0.6567,0.7818,0.2329); rgb(182pt)=(0.6701,0.7796,0.2239); rgb(183pt)=(0.6833,0.7773,0.2155); rgb(184pt)=(0.6963,0.775,0.2075); rgb(185pt)=(0.7091,0.7727,0.1998); rgb(186pt)=(0.7218,0.7703,0.1924); rgb(187pt)=(0.7344,0.7679,0.1852); rgb(188pt)=(0.7468,0.7654,0.1782); rgb(189pt)=(0.759,0.7629,0.1717); rgb(190pt)=(0.771,0.7604,0.1658); rgb(191pt)=(0.7829,0.7579,0.1608); rgb(192pt)=(0.7945,0.7554,0.157); rgb(193pt)=(0.806,0.7529,0.1546); rgb(194pt)=(0.8172,0.7505,0.1535); rgb(195pt)=(0.8281,0.7481,0.1536); rgb(196pt)=(0.8389,0.7457,0.1546); rgb(197pt)=(0.8495,0.7435,0.1564); rgb(198pt)=(0.86,0.7413,0.1587); rgb(199pt)=(0.8703,0.7392,0.1615); rgb(200pt)=(0.8804,0.7372,0.165); rgb(201pt)=(0.8903,0.7353,0.1695); rgb(202pt)=(0.9,0.7336,0.1749); rgb(203pt)=(0.9093,0.7321,0.1815); rgb(204pt)=(0.9184,0.7308,0.189); rgb(205pt)=(0.9272,0.7298,0.1973); rgb(206pt)=(0.9357,0.729,0.2061); rgb(207pt)=(0.944,0.7285,0.2151); rgb(208pt)=(0.9523,0.7284,0.2237); rgb(209pt)=(0.9606,0.7285,0.2312); rgb(210pt)=(0.9689,0.7292,0.2373); rgb(211pt)=(0.977,0.7304,0.2418); rgb(212pt)=(0.9842,0.733,0.2446); rgb(213pt)=(0.99,0.7365,0.2429); rgb(214pt)=(0.9946,0.7407,0.2394); rgb(215pt)=(0.9966,0.7458,0.2351); rgb(216pt)=(0.9971,0.7513,0.2309); rgb(217pt)=(0.9972,0.7569,0.2267); rgb(218pt)=(0.9971,0.7626,0.2224); rgb(219pt)=(0.9969,0.7683,0.2181); rgb(220pt)=(0.9966,0.774,0.2138); rgb(221pt)=(0.9962,0.7798,0.2095); rgb(222pt)=(0.9957,0.7856,0.2053); rgb(223pt)=(0.9949,0.7915,0.2012); rgb(224pt)=(0.9938,0.7974,0.1974); rgb(225pt)=(0.9923,0.8034,0.1939); rgb(226pt)=(0.9906,0.8095,0.1906); rgb(227pt)=(0.9885,0.8156,0.1875); rgb(228pt)=(0.9861,0.8218,0.1846); rgb(229pt)=(0.9835,0.828,0.1817); rgb(230pt)=(0.9807,0.8342,0.1787); rgb(231pt)=(0.9778,0.8404,0.1757); rgb(232pt)=(0.9748,0.8467,0.1726); rgb(233pt)=(0.972,0.8529,0.1695); rgb(234pt)=(0.9694,0.8591,0.1665); rgb(235pt)=(0.9671,0.8654,0.1636); rgb(236pt)=(0.9651,0.8716,0.1608); rgb(237pt)=(0.9634,0.8778,0.1582); rgb(238pt)=(0.9619,0.884,0.1557); rgb(239pt)=(0.9608,0.8902,0.1532); rgb(240pt)=(0.9601,0.8963,0.1507); rgb(241pt)=(0.9596,0.9023,0.148); rgb(242pt)=(0.9595,0.9084,0.145); rgb(243pt)=(0.9597,0.9143,0.1418); rgb(244pt)=(0.9601,0.9203,0.1382); rgb(245pt)=(0.9608,0.9262,0.1344); rgb(246pt)=(0.9618,0.932,0.1304); rgb(247pt)=(0.9629,0.9379,0.1261); rgb(248pt)=(0.9642,0.9437,0.1216); rgb(249pt)=(0.9657,0.9494,0.1168); rgb(250pt)=(0.9674,0.9552,0.1116); rgb(251pt)=(0.9692,0.9609,0.1061); rgb(252pt)=(0.9711,0.9667,0.1001); rgb(253pt)=(0.973,0.9724,0.0938); rgb(254pt)=(0.9749,0.9782,0.0872); rgb(255pt)=(0.9769,0.9839,0.0805)},
colorbar,
colorbar style={ytick={160,140,120}, yticklabels={\hspace{-0.17cm} -10,\hspace{-0.17cm} -30,\hspace{-0.17cm} -50}, style={xshift=-0.2cm}}
]
\addplot [forget plot] graphics [xmin=-15.0125, xmax=15.0125, ymin=0.9875, ymax=31.0125] {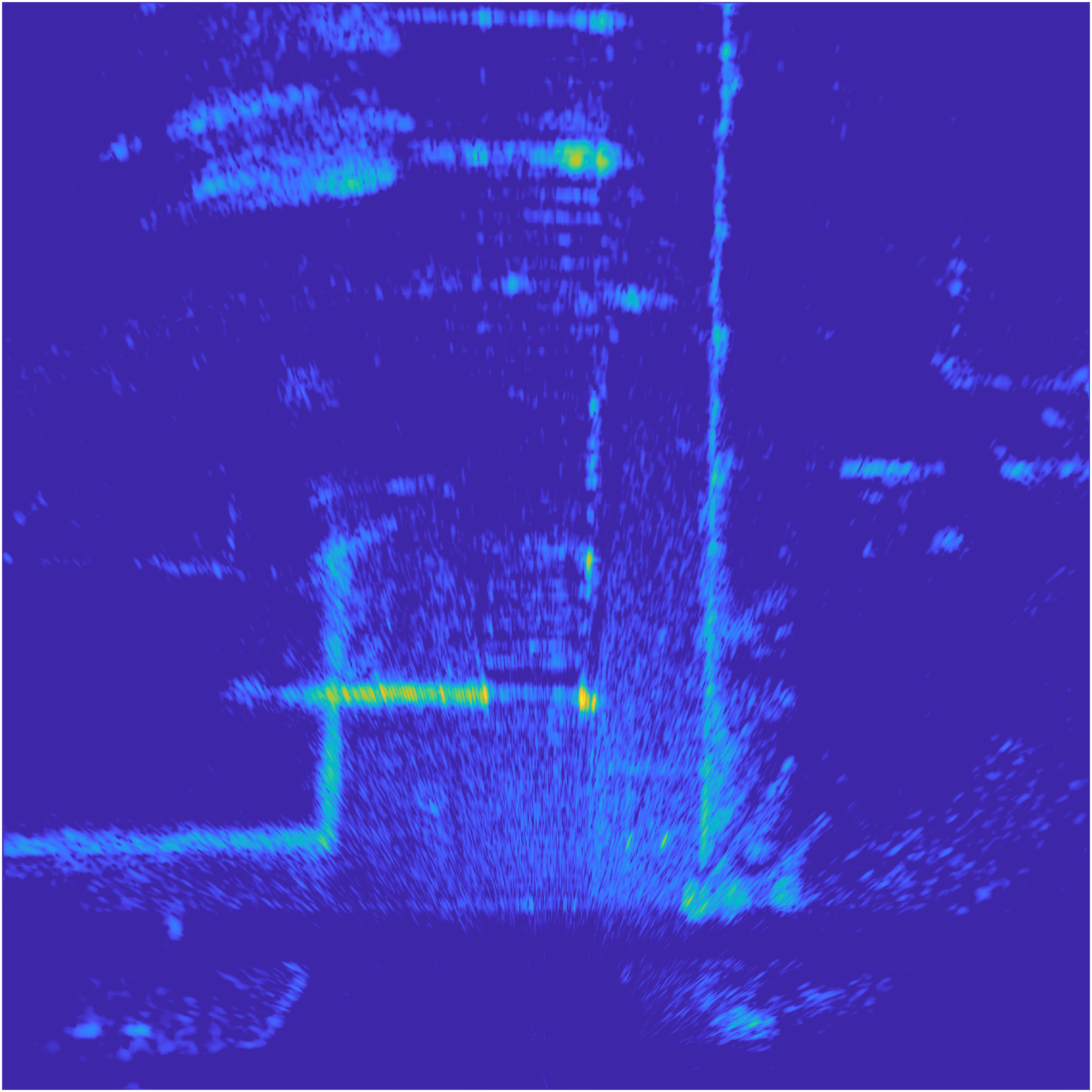};
\end{axis}
\end{tikzpicture}%
		\end{minipage}
	\end{subfigure}
	
	\begin{subfigure}{8.8cm}
		\begin{minipage}[l]{0.2cm}
			\caption{} \label{holo_diff}
		\end{minipage}
		\hfill
		\begin{minipage}[r]{8.1cm}
%
%
\begin{tikzpicture}

\begin{axis}[%
scale = 0.58, 
width=3.987in,
height=1.993in*12/15,
at={(0.669in,1.267in)},
scale only axis,
point meta min=-12,
point meta max=6,
axis on top,
xmin=-15,
xmax=15,
xlabel style={font=\color{white!15!black}, yshift = 0.2cm},
xlabel={x in m},
ymin=1,
ymax=13,
ylabel style={font=\color{white!15!black}, yshift = -0.6cm},
ylabel={y in m},
axis background/.style={fill=white},
title style={font=\bfseries},
colormap={mymap}{[1pt] rgb(0pt)=(0.2422,0.1504,0.6603); rgb(1pt)=(0.2444,0.1534,0.6728); rgb(2pt)=(0.2464,0.1569,0.6847); rgb(3pt)=(0.2484,0.1607,0.6961); rgb(4pt)=(0.2503,0.1648,0.7071); rgb(5pt)=(0.2522,0.1689,0.7179); rgb(6pt)=(0.254,0.1732,0.7286); rgb(7pt)=(0.2558,0.1773,0.7393); rgb(8pt)=(0.2576,0.1814,0.7501); rgb(9pt)=(0.2594,0.1854,0.761); rgb(11pt)=(0.2628,0.1932,0.7828); rgb(12pt)=(0.2645,0.1972,0.7937); rgb(13pt)=(0.2661,0.2011,0.8043); rgb(14pt)=(0.2676,0.2052,0.8148); rgb(15pt)=(0.2691,0.2094,0.8249); rgb(16pt)=(0.2704,0.2138,0.8346); rgb(17pt)=(0.2717,0.2184,0.8439); rgb(18pt)=(0.2729,0.2231,0.8528); rgb(19pt)=(0.274,0.228,0.8612); rgb(20pt)=(0.2749,0.233,0.8692); rgb(21pt)=(0.2758,0.2382,0.8767); rgb(22pt)=(0.2766,0.2435,0.884); rgb(23pt)=(0.2774,0.2489,0.8908); rgb(24pt)=(0.2781,0.2543,0.8973); rgb(25pt)=(0.2788,0.2598,0.9035); rgb(26pt)=(0.2794,0.2653,0.9094); rgb(27pt)=(0.2798,0.2708,0.915); rgb(28pt)=(0.2802,0.2764,0.9204); rgb(29pt)=(0.2806,0.2819,0.9255); rgb(30pt)=(0.2809,0.2875,0.9305); rgb(31pt)=(0.2811,0.293,0.9352); rgb(32pt)=(0.2813,0.2985,0.9397); rgb(33pt)=(0.2814,0.304,0.9441); rgb(34pt)=(0.2814,0.3095,0.9483); rgb(35pt)=(0.2813,0.315,0.9524); rgb(36pt)=(0.2811,0.3204,0.9563); rgb(37pt)=(0.2809,0.3259,0.96); rgb(38pt)=(0.2807,0.3313,0.9636); rgb(39pt)=(0.2803,0.3367,0.967); rgb(40pt)=(0.2798,0.3421,0.9702); rgb(41pt)=(0.2791,0.3475,0.9733); rgb(42pt)=(0.2784,0.3529,0.9763); rgb(43pt)=(0.2776,0.3583,0.9791); rgb(44pt)=(0.2766,0.3638,0.9817); rgb(45pt)=(0.2754,0.3693,0.984); rgb(46pt)=(0.2741,0.3748,0.9862); rgb(47pt)=(0.2726,0.3804,0.9881); rgb(48pt)=(0.271,0.386,0.9898); rgb(49pt)=(0.2691,0.3916,0.9912); rgb(50pt)=(0.267,0.3973,0.9924); rgb(51pt)=(0.2647,0.403,0.9935); rgb(52pt)=(0.2621,0.4088,0.9946); rgb(53pt)=(0.2591,0.4145,0.9955); rgb(54pt)=(0.2556,0.4203,0.9965); rgb(55pt)=(0.2517,0.4261,0.9974); rgb(56pt)=(0.2473,0.4319,0.9983); rgb(57pt)=(0.2424,0.4378,0.9991); rgb(58pt)=(0.2369,0.4437,0.9996); rgb(59pt)=(0.2311,0.4497,0.9995); rgb(60pt)=(0.225,0.4559,0.9985); rgb(61pt)=(0.2189,0.462,0.9968); rgb(62pt)=(0.2128,0.4682,0.9948); rgb(63pt)=(0.2066,0.4743,0.9926); rgb(64pt)=(0.2006,0.4803,0.9906); rgb(65pt)=(0.195,0.4861,0.9887); rgb(66pt)=(0.1903,0.4919,0.9867); rgb(67pt)=(0.1869,0.4975,0.9844); rgb(68pt)=(0.1847,0.503,0.9819); rgb(69pt)=(0.1831,0.5084,0.9793); rgb(70pt)=(0.1818,0.5138,0.9766); rgb(71pt)=(0.1806,0.5191,0.9738); rgb(72pt)=(0.1795,0.5244,0.9709); rgb(73pt)=(0.1785,0.5296,0.9677); rgb(74pt)=(0.1778,0.5349,0.9641); rgb(75pt)=(0.1773,0.5401,0.9602); rgb(76pt)=(0.1768,0.5452,0.956); rgb(77pt)=(0.1764,0.5504,0.9516); rgb(78pt)=(0.1755,0.5554,0.9473); rgb(79pt)=(0.174,0.5605,0.9432); rgb(80pt)=(0.1716,0.5655,0.9393); rgb(81pt)=(0.1686,0.5705,0.9357); rgb(82pt)=(0.1649,0.5755,0.9323); rgb(83pt)=(0.161,0.5805,0.9289); rgb(84pt)=(0.1573,0.5854,0.9254); rgb(85pt)=(0.154,0.5902,0.9218); rgb(86pt)=(0.1513,0.595,0.9182); rgb(87pt)=(0.1492,0.5997,0.9147); rgb(88pt)=(0.1475,0.6043,0.9113); rgb(89pt)=(0.1461,0.6089,0.908); rgb(90pt)=(0.1446,0.6135,0.905); rgb(91pt)=(0.1429,0.618,0.9022); rgb(92pt)=(0.1408,0.6226,0.8998); rgb(93pt)=(0.1383,0.6272,0.8975); rgb(94pt)=(0.1354,0.6317,0.8953); rgb(95pt)=(0.1321,0.6363,0.8932); rgb(96pt)=(0.1288,0.6408,0.891); rgb(97pt)=(0.1253,0.6453,0.8887); rgb(98pt)=(0.1219,0.6497,0.8862); rgb(99pt)=(0.1185,0.6541,0.8834); rgb(100pt)=(0.1152,0.6584,0.8804); rgb(101pt)=(0.1119,0.6627,0.877); rgb(102pt)=(0.1085,0.6669,0.8734); rgb(103pt)=(0.1048,0.671,0.8695); rgb(104pt)=(0.1009,0.675,0.8653); rgb(105pt)=(0.0964,0.6789,0.8609); rgb(106pt)=(0.0914,0.6828,0.8562); rgb(107pt)=(0.0855,0.6865,0.8513); rgb(108pt)=(0.0789,0.6902,0.8462); rgb(109pt)=(0.0713,0.6938,0.8409); rgb(110pt)=(0.0628,0.6972,0.8355); rgb(111pt)=(0.0535,0.7006,0.8299); rgb(112pt)=(0.0433,0.7039,0.8242); rgb(113pt)=(0.0328,0.7071,0.8183); rgb(114pt)=(0.0234,0.7103,0.8124); rgb(115pt)=(0.0155,0.7133,0.8064); rgb(116pt)=(0.0091,0.7163,0.8003); rgb(117pt)=(0.0046,0.7192,0.7941); rgb(118pt)=(0.0019,0.722,0.7878); rgb(119pt)=(0.0009,0.7248,0.7815); rgb(120pt)=(0.0018,0.7275,0.7752); rgb(121pt)=(0.0046,0.7301,0.7688); rgb(122pt)=(0.0094,0.7327,0.7623); rgb(123pt)=(0.0162,0.7352,0.7558); rgb(124pt)=(0.0253,0.7376,0.7492); rgb(125pt)=(0.0369,0.74,0.7426); rgb(126pt)=(0.0504,0.7423,0.7359); rgb(127pt)=(0.0638,0.7446,0.7292); rgb(128pt)=(0.077,0.7468,0.7224); rgb(129pt)=(0.0899,0.7489,0.7156); rgb(130pt)=(0.1023,0.751,0.7088); rgb(131pt)=(0.1141,0.7531,0.7019); rgb(132pt)=(0.1252,0.7552,0.695); rgb(133pt)=(0.1354,0.7572,0.6881); rgb(134pt)=(0.1448,0.7593,0.6812); rgb(135pt)=(0.1532,0.7614,0.6741); rgb(136pt)=(0.1609,0.7635,0.6671); rgb(137pt)=(0.1678,0.7656,0.6599); rgb(138pt)=(0.1741,0.7678,0.6527); rgb(139pt)=(0.1799,0.7699,0.6454); rgb(140pt)=(0.1853,0.7721,0.6379); rgb(141pt)=(0.1905,0.7743,0.6303); rgb(142pt)=(0.1954,0.7765,0.6225); rgb(143pt)=(0.2003,0.7787,0.6146); rgb(144pt)=(0.2061,0.7808,0.6065); rgb(145pt)=(0.2118,0.7828,0.5983); rgb(146pt)=(0.2178,0.7849,0.5899); rgb(147pt)=(0.2244,0.7869,0.5813); rgb(148pt)=(0.2318,0.7887,0.5725); rgb(149pt)=(0.2401,0.7905,0.5636); rgb(150pt)=(0.2491,0.7922,0.5546); rgb(151pt)=(0.2589,0.7937,0.5454); rgb(152pt)=(0.2695,0.7951,0.536); rgb(153pt)=(0.2809,0.7964,0.5266); rgb(154pt)=(0.2929,0.7975,0.517); rgb(155pt)=(0.3052,0.7985,0.5074); rgb(156pt)=(0.3176,0.7994,0.4975); rgb(157pt)=(0.3301,0.8002,0.4876); rgb(158pt)=(0.3424,0.8009,0.4774); rgb(159pt)=(0.3548,0.8016,0.4669); rgb(160pt)=(0.3671,0.8021,0.4563); rgb(161pt)=(0.3795,0.8026,0.4454); rgb(162pt)=(0.3921,0.8029,0.4344); rgb(163pt)=(0.405,0.8031,0.4233); rgb(164pt)=(0.4184,0.803,0.4122); rgb(165pt)=(0.4322,0.8028,0.4013); rgb(166pt)=(0.4463,0.8024,0.3904); rgb(167pt)=(0.4608,0.8018,0.3797); rgb(168pt)=(0.4753,0.8011,0.3691); rgb(169pt)=(0.4899,0.8002,0.3586); rgb(170pt)=(0.5044,0.7993,0.348); rgb(171pt)=(0.5187,0.7982,0.3374); rgb(172pt)=(0.5329,0.797,0.3267); rgb(173pt)=(0.547,0.7957,0.3159); rgb(175pt)=(0.5748,0.7929,0.2941); rgb(176pt)=(0.5886,0.7913,0.2833); rgb(177pt)=(0.6024,0.7896,0.2726); rgb(178pt)=(0.6161,0.7878,0.2622); rgb(179pt)=(0.6297,0.7859,0.2521); rgb(180pt)=(0.6433,0.7839,0.2423); rgb(181pt)=(0.6567,0.7818,0.2329); rgb(182pt)=(0.6701,0.7796,0.2239); rgb(183pt)=(0.6833,0.7773,0.2155); rgb(184pt)=(0.6963,0.775,0.2075); rgb(185pt)=(0.7091,0.7727,0.1998); rgb(186pt)=(0.7218,0.7703,0.1924); rgb(187pt)=(0.7344,0.7679,0.1852); rgb(188pt)=(0.7468,0.7654,0.1782); rgb(189pt)=(0.759,0.7629,0.1717); rgb(190pt)=(0.771,0.7604,0.1658); rgb(191pt)=(0.7829,0.7579,0.1608); rgb(192pt)=(0.7945,0.7554,0.157); rgb(193pt)=(0.806,0.7529,0.1546); rgb(194pt)=(0.8172,0.7505,0.1535); rgb(195pt)=(0.8281,0.7481,0.1536); rgb(196pt)=(0.8389,0.7457,0.1546); rgb(197pt)=(0.8495,0.7435,0.1564); rgb(198pt)=(0.86,0.7413,0.1587); rgb(199pt)=(0.8703,0.7392,0.1615); rgb(200pt)=(0.8804,0.7372,0.165); rgb(201pt)=(0.8903,0.7353,0.1695); rgb(202pt)=(0.9,0.7336,0.1749); rgb(203pt)=(0.9093,0.7321,0.1815); rgb(204pt)=(0.9184,0.7308,0.189); rgb(205pt)=(0.9272,0.7298,0.1973); rgb(206pt)=(0.9357,0.729,0.2061); rgb(207pt)=(0.944,0.7285,0.2151); rgb(208pt)=(0.9523,0.7284,0.2237); rgb(209pt)=(0.9606,0.7285,0.2312); rgb(210pt)=(0.9689,0.7292,0.2373); rgb(211pt)=(0.977,0.7304,0.2418); rgb(212pt)=(0.9842,0.733,0.2446); rgb(213pt)=(0.99,0.7365,0.2429); rgb(214pt)=(0.9946,0.7407,0.2394); rgb(215pt)=(0.9966,0.7458,0.2351); rgb(216pt)=(0.9971,0.7513,0.2309); rgb(217pt)=(0.9972,0.7569,0.2267); rgb(218pt)=(0.9971,0.7626,0.2224); rgb(219pt)=(0.9969,0.7683,0.2181); rgb(220pt)=(0.9966,0.774,0.2138); rgb(221pt)=(0.9962,0.7798,0.2095); rgb(222pt)=(0.9957,0.7856,0.2053); rgb(223pt)=(0.9949,0.7915,0.2012); rgb(224pt)=(0.9938,0.7974,0.1974); rgb(225pt)=(0.9923,0.8034,0.1939); rgb(226pt)=(0.9906,0.8095,0.1906); rgb(227pt)=(0.9885,0.8156,0.1875); rgb(228pt)=(0.9861,0.8218,0.1846); rgb(229pt)=(0.9835,0.828,0.1817); rgb(230pt)=(0.9807,0.8342,0.1787); rgb(231pt)=(0.9778,0.8404,0.1757); rgb(232pt)=(0.9748,0.8467,0.1726); rgb(233pt)=(0.972,0.8529,0.1695); rgb(234pt)=(0.9694,0.8591,0.1665); rgb(235pt)=(0.9671,0.8654,0.1636); rgb(236pt)=(0.9651,0.8716,0.1608); rgb(237pt)=(0.9634,0.8778,0.1582); rgb(238pt)=(0.9619,0.884,0.1557); rgb(239pt)=(0.9608,0.8902,0.1532); rgb(240pt)=(0.9601,0.8963,0.1507); rgb(241pt)=(0.9596,0.9023,0.148); rgb(242pt)=(0.9595,0.9084,0.145); rgb(243pt)=(0.9597,0.9143,0.1418); rgb(244pt)=(0.9601,0.9203,0.1382); rgb(245pt)=(0.9608,0.9262,0.1344); rgb(246pt)=(0.9618,0.932,0.1304); rgb(247pt)=(0.9629,0.9379,0.1261); rgb(248pt)=(0.9642,0.9437,0.1216); rgb(249pt)=(0.9657,0.9494,0.1168); rgb(250pt)=(0.9674,0.9552,0.1116); rgb(251pt)=(0.9692,0.9609,0.1061); rgb(252pt)=(0.9711,0.9667,0.1001); rgb(253pt)=(0.973,0.9724,0.0938); rgb(254pt)=(0.9749,0.9782,0.0872); rgb(255pt)=(0.9769,0.9839,0.0805)},
colorbar,
colorbar style={ytick={5,0,-5,-10}, yticklabels={\hspace{-0.17cm} 5,\hspace{-0.17cm} 0,\hspace{-0.17cm} -5, \hspace{-0.17cm} -10}, style={xshift=-0.2cm}}
]
\addplot [forget plot] graphics [xmin=-15.0125, xmax=15.0125, ymin=0.9875, ymax=31.0125] {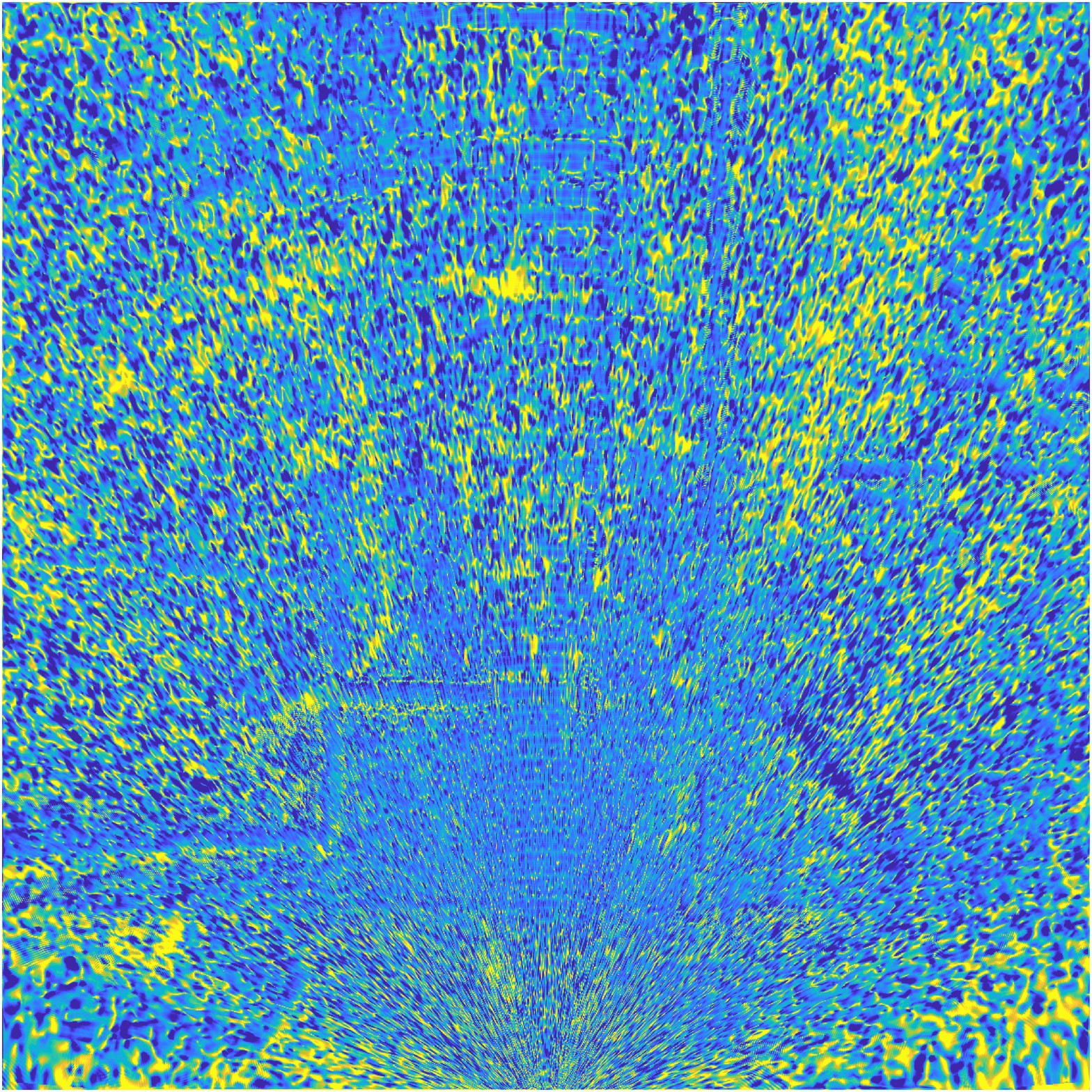};
\end{axis}
\end{tikzpicture}%
		\end{minipage}
	\end{subfigure}
	
	\caption{a) Result of the optimized BP in dB normalized to its maximum amplitude; b) amplitude difference to the reference image in dB.}

	\label{fig:kombination_neu}
\end{figure}

The resulting image in Fig. \ref{holo_kombi} was interpolated into the original grid for comparability.
Fig. \ref{holo_diff} shows the difference from the reference image in Fig. \ref{fig:holo_ref}, which was mostly in the range of \SI{-6}{\dB} due to the reduced number of antennas.

\begin{table}
	\centering
	Algorithm 2. \> Optimized Back-Projection algorithm.\\[0.7em]
	\setlength{\tabcolsep}{0.1pt}
	\begin{tabular}{C{0.3cm}L{0.3cm}L{0.3cm}L{0.3cm}L{0.3cm}L{0.3cm}|}
		{\fontsize{6}{9}\selectfont1}& \multicolumn{5}{l}{\textbf{procedure} CALCULATE SAR-IMAGE($\vec{p}$, $\vec{q}_{\text{tx}}$, $\vec{q}_{\text{rx}}$, $f_{\text{doppler}}$, $s$, $w_{\text{sar}}$)} \\
		{\fontsize{6}{9}\selectfont2}& & \multicolumn{4}{l}{\sethlcolor{soulred}\hl{\textbf{for all} pixels $p$ \textbf{do}}} \\
		
		{\fontsize{6}{9}\selectfont3}& & & \multicolumn{3}{l}{\textbf{for all} measurements $m$ \textbf{do}} \\
		{\fontsize{6}{9}\selectfont4}& & & & \multicolumn{2}{l}{$d_{\text{tx}}(p,m) \gets || \vec{p}(p) - \vec{q}_{\text{tx}}(m)||_2$} \\
		
		{\fontsize{6}{9}\selectfont5}& & &\multicolumn{1}{l}{} & \multicolumn{2}{l}{\sethlcolor{soulblue}\hl{\textbf{for all} rx antennas $n_{\text{rx}}$ \textbf{do}} }\\
	
		{\fontsize{6}{9}\selectfont6}& & & & & \multicolumn{1}{l}{$d_{\text{rx}}(p,m,n_{\text{rx}}) \gets || \vec{p}(p) - \vec{q}_{\text{rx}}(m,n_{\text{rx}}) ||_2$} \\
		
		{\fontsize{6}{9}\selectfont7}& & & & & \multicolumn{1}{l}{$d_{\text{hyp}}(p,m,n_{\text{rx}}) \gets d_{\text{tx}}(p,m) + d_{\text{rx}}(p,m,n_{\text{rx}})$} \\
		
		{\fontsize{6}{9}\selectfont8}& & & & & \multicolumn{1}{l}{\sethlcolor{soulmagenta}\hl{$f_{\text{ind}}(p,m,n_{\text{rx}})$}$ \gets $\sethlcolor{soulmagenta}\hl{$a_{\text{1}}$}$ \cdot d_{\text{hyp}}(p,m,n_{\text{rx}}) + $\sethlcolor{soulgreen}\hl{$f_{\text{doppler}}(p) $}} \\
		
		{\fontsize{6}{9}\selectfont9}& & & & & \multicolumn{1}{l}{$s_{\text{hyp}}(p,m,n_{\text{rx}}) \gets \exp($\sethlcolor{soulmagenta}\hl{$a_{\text{2}}$}$ \cdot d_{\text{hyp}}(p,m,n_{\text{rx}}))$} \\

		{\fontsize{6}{9}\selectfont10}& & & & & \multicolumn{1}{l}{$P(p,m,n_{\text{rx}}) \gets s_{\text{hyp}}(p,m,n_{\text{rx}}) \cdot $\sethlcolor{soulmagenta}\hl{$s(f_{\text{ind}},m,n_{\text{rx}})$} $\cdot$ \sethlcolor{soulviolet}\hl{$w_{sar}$}} \\
		{\fontsize{6}{9}\selectfont11}& & \multicolumn{4}{l}{\textbf{for all} pixels $p$ \textbf{do}} \\
		{\fontsize{6}{9}\selectfont12}& & & \multicolumn{3}{l}{$P(p) \gets \sum_{m}^{} \sum_{n_{\text{rx}}}^{}  P(p,m,n_{\text{rx}})$} \\
		
	\end{tabular}
\end{table}

\section{Conclusion}

In this work, the Back-Projection algorithm was optimized to enable real-time processing in automotive SAR imaging.
Six different measures were analyzed individually with respect to their influence on image quality before they were implemented in combination.
The real-time capability could be confirmed in tests on two different GPUs by reconstructing a \qtyproduct{30 x 30}{\m} area within \SI{18.8}{\ms} and \SI{28.4}{\ms}, respectively, without any noticeable loss of quality.
Since this was four to five times faster than the actual radar measurement time, even a continuous SAR measurement could be processed without increasing delay.
Also, the amount of processed data was drastically reduced, allowing GPUs to be used with fewer resources and a lower power consumption.
These investigations therefore present a catalog of measures that operators can individually adapt to their radar and processing systems with their corresponding FMCW parameters, resolution, SNR, and number of antennas.
In the future, we will adapt this approach to an embedded system to further reduce the demands on power consumption and processing resources. 

\section*{Acknowledgment}

The authors would like to thank the Symeo team from indie Semiconductor (Mark Christmann and Jannis Groh) for their support with the measurement setup and radar system.

\end{document}